\documentclass[aps,showpacs,a4paper,floatfix,twocolumn,prc,amsmath,amssymb]{revtex4-1}
\usepackage{graphicx}
\usepackage{epsfig}
\usepackage{ulem}
\usepackage{morefloats}
\usepackage{rotating}
\usepackage{xcolor}
\usepackage{soul}

\newcommand{\pc}[1]{\ensuremath{\left(#1\right)}}

\begin{document}

\title{ An Evaluation Of  The Possible Effect Of Tetra-neutron Production On Ternary Fission Yields}

\author{H. Pais}  
\email{hpais@uc.pt}
\affiliation{CFisUC, Department of Physics, University of Coimbra, 3004-516 Coimbra, Portugal.}
\author{G. R\"opke}
\email{gerd.roepke@uni-rostock.de}
\affiliation{Institut f\"ur Physik, Universit\"at Rostock, D-18051 Rostock, Germany.}
\author{J. B. Natowitz}
\email{natowitz@comp.tamu.edu}
\affiliation{Cyclotron Institute, Texas A\&M University, College Station, Texas 77843, USA.}

 \begin{abstract}

  In this work, we study the effect of including a resonant state of four neutrons in the low-density warm nuclear equation of state, using a relativistic mean-field formalism, where in-medium effects are considered. 
  For that purpose, the abundances of 62 different clusters immersed in a gas of protons and neutrons, are calculated with and without the presence of this resonant tetraneutron state. 
Ternary fission experiments are environments where not only these clusters can be formed, but also where such thermodynamic conditions of low densities and moderate temperatures can be achieved.
 The calculated yields of isotopes with and without the tetraneutron included are then compared with experimentally observed ternary fission yields. 
 While the results indicate that the tetraneutron has little effect on yields of other clusters under the existing ternary fission conditions, a possibly observable effect on yields of clusters emitted from the neck region in mid-peripheral heavy ion collisions is clearly suggested.

\end{abstract}

\maketitle

\section{\label{intro} Introduction}

The possible existence of tetra-neutrons has long been postulated and searches for both stable and resonance state tetra-neutrons continue to be pursued. See, for example, Refs.~\cite{Fujioka23,Muzalevskii23,Marques2021} and references therein. The first report of a possible $4n$ bound state was presented in 2002 by Marqu\'es {\it et al.}~\cite{Marques2002} based on a reaction $^{14}$Be$\rightarrow ^{10}$Be$+4n$. Later, Marqu\'es {\it et al.}~\cite{Marques2005}  reported a resonant state with an energy of $\sim 2$ MeV.  

The current most credible evidence of tetra-neutron formation appears to be that of Duer et al. \cite{Duer22}, based upon  an experiment at RIKEN with the SAMURAI spectrometer, using a high-energy beam of $^8$He on a proton target. Evidence for a resonant four neutron state with an energy of $E_{4n} = 2.37 \pm 0.38$(stat) $\pm 0.44$(sys) MeV and a width of $\Gamma = 1.75 \pm 0.22$(stat) $\pm 0.30$(sys) MeV was reported. This value for the energy is substantially higher (and the width lower) than found in a previous experiment \citep{Kisamori2016}, where a high-energy radioactive $^8$He beam incident on a liquid $^4$He target, producing an apparent resonant tetraneutron state, with an energy of $E_{4n}=0.8 \pm 1.4$ MeV, with $\Gamma=2.6$ MeV as an upper limit.
The role that a tetra-neutron state could play in neutron star matter has been addressed in Pais et al \cite{Pais23}. 

In Ref.~\cite{Pais23} the authors studied the effect of the tetraneutron on the equation of state with light clusters at finite temperature and fixed proton fraction, for thermodynamic conditions observed in heavy-ion collisions and in core-collapse supernovae. They found that the systems most prone to manifest the effect of such exotic clusters would be neutron stars, due to their very-low temperature and very neutron-rich environment. In this work, we analyze the effect the existence of $4n$ clusters would have on mass fractions of other light clusters in the matter considered.  Whereas in \cite{Pais23} only $^{2,3}$H and
$^{3,4,6}$He were included in the calculation, here we include a total of 62 light clusters (isotopes of the elements with $Z=1-6$). We consider different temperatures and small proton fractions. The complete list of the clusters included, together with their properties, is given in Table IX and X of Ref.~\cite{Natowitz23}. 
For the present calculations we adopt the Duer {\it et al.} $4n$ binding energy. We note that it was reported in \cite{Pais23} that the difference in the abundances of the light charged clusters was small when two different ranges for the binding energy of $4n$ were considered.

For an initial  baseline test of the theoretical calculations we compare cluster yields, obtained with and without the inclusion of the tetraneutron resonance, with previously reported experimental yields observed in ternary fission experiments \cite{Natowitz23}. Ternary ﬁssion of actinides occurs in the neck region between the two heaviest fission fragments and probes the state of the nucleus at scission. In a previous publication \cite{Natowitz23} we explored the ternary fission in the $^{241}$Pu($n_{th},f$) reaction. Based upon the comparisons of theoretical yields to experimental yields we concluded that the neck region could be characterized by a temperature of 1.29 MeV, a density of $6.7 \times 10^{-5}$ fm$^{-3}$ and a proton fraction 0.035. Similar parameter values are found in other ternary fission processes \cite{Roepke21}. Given the extreme neutron richness found for the ternary fragment region in fission it is natural to ask whether the production of tetraneutrons might exist and be reflected in these ternary fragment yields.

The structure of the article is as follows: in Section 2, we briefly describe the RMF formalism, and in Section 3, we present the results. Finally, we draw conclusions in Section 4.

\section{Formalism}

We use a relativistic mean-field formalism, that includes as degrees of freedom nucleons and clusters, and considers in-medium effects between the clusters and the medium. These clusters include the H, He, Li, Be, B and C isotopes. Following Ref.~\cite{Pais23}, we also include the tetraneutron, in the same spirit of the other bosonic clusters.

The interaction is mediated by the exchange of virtual mesons, namely the scalar-isoscalar $\phi$ meson with mass $m_s$, the vector-isoscalar $\omega$ meson with mass $m_v$, and the isovector-vector $\rho$ meson with mass $m_\rho$. 
The Lagrangian density is given by \citep{typel10,ferreira12,pais15,avancini17,Pais18,Pais2019}
\begin{eqnarray}
{\cal L} &=& \sum_{j=n,p,cl} {\cal L}_{j}                            
+ {\cal L}_{\sigma} + {\cal L}_{\omega} + {\cal L}_{\rho}+ {\cal L}_{\omega\rho}.
\end{eqnarray}
For the fermionic clusters,  $f$, with half-integer spin, we have
\begin{eqnarray}
{\cal L}_f &=& \bar{\psi}\left[\gamma_\mu i D_f^\mu - M_f^*\right]\psi,
\end{eqnarray}
with  
\begin{equation}
iD^{\mu }_f = i \partial ^{\mu }-g_{vf} \omega^{\mu }-\frac{g_\rho}{2}{\boldsymbol\tau}_f \cdot \mathbf{b}^\mu ,
\end{equation}
where ${\boldsymbol \tau}_f$ are the Pauli matrices and $g_{vf}$ is the
  coupling of cluster $f$ to the vector meson $\omega$, which is defined
  as $g_{vf}=A_f g_v$ for all clusters.

The Lagrangian density for the bosonic clusters, $b$, including $4n$, with integer spin, is given by
\begin{eqnarray}
\mathcal{L}_{b,s=0}&=&\frac{1}{2} (i D^{\mu}_{b} \phi_{b})^*
(i D_{\mu b} \phi_{b})-\frac{1}{2}\phi_{b}^* \pc{M_{b}^*}^2
\phi_{b},\\
\mathcal{L}_{b,s\neq0}&=&\frac{1}{4} (i D^{\mu}_{b} \phi^{\nu}_{b}-
i D^{\nu}_{b} \phi^{\mu}_{b})^*
(i D_{b\mu} \phi_{b\nu}-i D_{b\nu} \phi_{b\mu})\nonumber\\
&&-\frac{1}{2}\phi^{\mu *}_{b} \pc{M_{b}^*}^2 \phi_{b\mu},
\end{eqnarray}
with
\begin{equation}
iD^{\mu }_b = i \partial ^{\mu }-g_{vb} \omega^{\mu } \, ,
\end{equation}
where $s$ is the spin of the cluster $b$.

For the nucleonic gas,  $j=n,p$, we have
\begin{eqnarray}
{\cal L}_j &=& \bar{\psi}\left[\gamma_\mu i D^\mu - m^*\right]\psi
,\end{eqnarray}
with
\begin{eqnarray}
i D^\mu&=&i\partial^\mu-g_v\omega^\mu-\frac{g_\rho}{2}{\boldsymbol\tau}_j \cdot \mathbf{b}^\mu \\
m^*&=&m-g_s\phi_0 \label{meff}
.\end{eqnarray}

For the contribution of the mesonic fields to the Lagrangian, we have the standard RMF expressions:
\begin{eqnarray}
{\cal L}_\sigma&=&+\frac{1}{2}\left(\partial_{\mu}\phi\partial^{\mu}\phi
-m_s^2 \phi^2 - \frac{1}{3}\kappa \phi^3 -\frac{1}{12}\lambda\phi^4\right),\nonumber\\
{\cal L}_\omega&=&-\frac{1}{4}\Omega_{\mu\nu}\Omega^{\mu\nu}+\frac{1}{2}
m_v^2 V_{\mu}V^{\mu}, \nonumber \\ 
{\cal L}_\rho&=&-\frac{1}{4}\mathbf B_{\mu\nu}\cdot\mathbf B^{\mu\nu}+\frac{1}{2}
m_\rho^2 \mathbf b_{\mu}\cdot \mathbf b^{\mu}, \nonumber \\ 
{\cal L}_{\omega\rho}&=& g_{\omega\rho} g_\rho^2 g_v^2 V_{\mu}V^{\mu}\mathbf b_{\nu}\cdot \mathbf b^{\nu}
,\end{eqnarray}
where
$\Omega_{\mu\nu}=\partial_{\mu}V_{\nu}-\partial_{\nu}V_{\mu} $ and $ \mathbf B_{\mu\nu}=\partial_{\mu}\mathbf b_{\nu}-\partial_{\nu} \mathbf b_{\mu}
- g_\rho (\mathbf b_\mu \times \mathbf b_\nu)$. The non-linear term ${\cal L}_{\omega\rho}$ mixes the vector sector to model the density dependence of the symmetry energy.

The total binding energy of a light cluster $cl$ is given by 
\begin{eqnarray}
B_{cl}=A_{cl} m^*-M_{cl}^* \,,  \label{binding} 
\end{eqnarray}
with $M_{cl}^*$ its effective mass of cluster, which is determined by
the scalar meson coupling  as well as by a binding energy shift: 
\begin{eqnarray}
M_{cl}^*=A_{cl} m - g_{scl}\phi_0 - \left(B_{cl}^0 + \delta B_{cl}\right) \, .
\label{meffi2}
\end{eqnarray}

In expression (\ref{meffi2}), $B^0_{cl}$ is the binding energy of the cluster $cl$ in the vacuum, and these constants are fixed to experimental values. For the tetraneutron, we take the values of \cite{Duer22}, and we take this value negative since this cluster was considered a resonant state. The other clusters take positive binding energy. 

We include a binding energy shift term, $\delta B_{cl}$, to include the effect of the medium. This term  is given in \citep{Pais18}, and takes the form
\begin{eqnarray}
\delta B_{cl}=\frac{Z_{cl}}{\rho_0}\left(\epsilon_p^*-m\rho_p^*\right)+\frac{N_{cl}}{\rho_0}\left(\epsilon_n^*-m\rho_n^*\right) \, .
\label{deltaB}
\end{eqnarray}
 This term acts as the energetic counterpart of the excluded volume mechanism in the Thomas-Fermi approximation or the Pauli blocking term in a quantum statistical approach \cite{typel10}. $\rho_0$ is the nuclear saturation density, $N_{cl}$ and $Z_{cl}$ are the neutron and proton numbers, and $\epsilon_j^*$ and $\rho_j^*$  are the energy density and density of the lowest energy levels of the  gas, respectively. In this way, the  energy states occupied by the gas are removed from the cluster binding energy, avoiding double-counting.

Regarding the scalar and vector cluster--meson couplings, we follow the prescription introduced in \cite{Pais18}. The scalar coupling is given by
\begin{equation}
\label{gs}
g_{scl}=x_{s} A_{cl} g_s, 
\end{equation}
 while the vector coupling is given by
\begin{equation}
\label{gv}
g_{vcl}=A_{cl} g_v \, ,
\end{equation}
with $A_{cl}$ the number of nucleons in each cluster. The $x_s$ factor can vary from 0 to 1. In a previous work, its value was fixed to $x_s=0.85\pm0.05$ from a fit to the Virial EoS \cite{Pais18}. In later works, the value was found to be higher, $x_s=0.92\pm0.02$, when a fit to experimental data was considered \citep{PaisPRL,PaisJPG}. 
The dissolution of the clusters is affected by a combination of both the binding energy shift, $\delta B_j$, and this factor $x_s$. Substituting eqs.~(\ref{meffi2}), (\ref{meff}), and (\ref{gs}) into eq.~(\ref{binding}), we obtain
\begin{eqnarray}
B_j&=&A_j g_s \phi_0 \left(x_{sj}-1 \right) + B_j^0 + \delta B_j \, .
\end{eqnarray}

This implies that a larger $x_{s}$ corresponds to a larger binding energy of the cluster,  meaning that the dissolution of the cluster will occur at larger densities.  
If $x_{s}=1$, the dissolution is totally defined by the binding shift $\delta B_{cl}$. In a previous work \cite{Pais18}, where we introduced these in-medium effects via the scalar cluster-meson coupling factor $x_s$ and the binding energy shift term $\delta B_{cl}$, we observed  that at finite temperature, the clusters  dissolved at a density well above that for which $B_{cl}\sim 0$ (see Fig.~2 of Ref.~\cite{Pais18}). 
The binding energy shift modifies the meson equations, so that the mass fractions of the clusters are affected in a self-consistent way.

\section{Results}
\label{sec:results}

In this section, we explore cluster mass fractions and yields
calculated as a function of density, with and without inclusion of the
tetraneutron, for the clusters for atomic number $Z=1-6$ and discuss
how they are affected by temperature and isospin asymmetry of the
medium. The choice of these isotopes is based on our previous analyses
of ternary fission yields. There, a suppression of yields for higher $Z$
isotopes was observed and attributed to nucleation time and/or finite
size effects \cite{Natowitz23,Sara}. 

\begin{figure}
 \begin{tabular}{c}
 \includegraphics[width=0.45\textwidth]{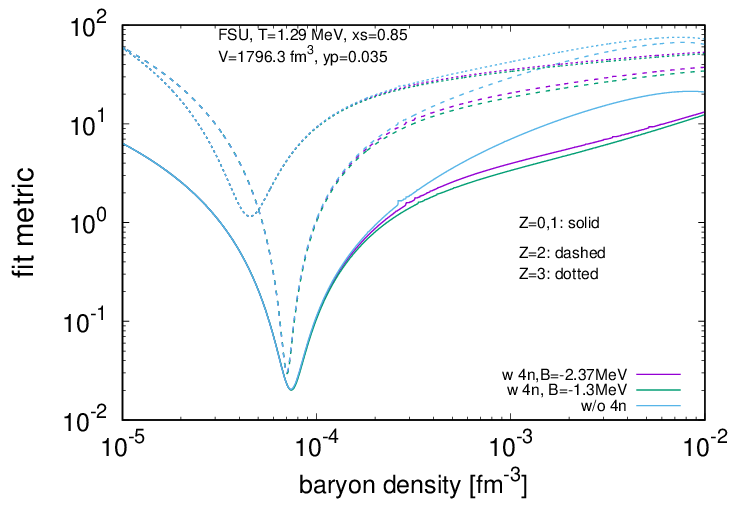} 
   \end{tabular}
  \caption{Fit metric as a function of the density for $T=1.29$ MeV, $y_p=0.035$.  The RMF FSU model was used with $x_s=0.85$. The results are shown for the isotopes of the elements with $Z=1-3$, where the calculation for  $Z=1$ also includes the free neutrons. The calculations are carried out with and without the $4n$ resonance, and considering two values for the binding energies. } 
\label{fig1} 
\end{figure}

Even though there is an initial  total of 62 cluster states many decay to others. As a result, we present  results for 25 different observable states. These ``effective densities'' $\widetilde{\rho_i}$, are the final calculated densities including such decays. The paths to these effective densities  are given below, but, for simplicity, we omit the symbols $\widetilde{\rho}$ and $\rho$:

\begin{eqnarray}
^3 {\rm H} \rightarrow ^3{\rm H}+^4{\rm H}^*+^6{\rm Li}^* \nonumber \\
^4{\rm He} \rightarrow ^4{\rm He}+^5{\rm He}+^6{\rm He}^*+^8{\rm He}^*+^8{\rm Be}+^9{\rm Be}^* \nonumber \\
^6{\rm He} \rightarrow ^6{\rm He}+^7{\rm He}^* \nonumber \\
^8{\rm He} \rightarrow ^8{\rm He}+^9{\rm He}^* \nonumber \\
^7{\rm Li} \rightarrow ^7{\rm Li}+^8{\rm Li}^* \nonumber \\
^9{\rm Li} \rightarrow ^9{\rm Li}+^{10}{\rm Li}^* \nonumber \\
^{11}{\rm Li} \rightarrow ^{11}{\rm Li}+^{12}{\rm Li}^* \nonumber \\
^{10}{\rm Be} \rightarrow ^{10}{\rm Be}+^{10}{\rm Be}^0+^{11}{\rm Be}^* \nonumber \\
^{11}{\rm Be} \rightarrow ^{11}{\rm Be}+^{11}{\rm Be}^0 \nonumber \\
^{12}{\rm Be} \rightarrow ^{12}{\rm Be}+^{12}{\rm Be}^0+^{13}{\rm Be}^* \nonumber \\
^{14}{\rm Be} \rightarrow ^{14}{\rm Be}+^{15}{\rm Be}^* \nonumber \\
^{10}{\rm B} \rightarrow ^{10}{\rm B}+^{10}{\rm B}^0 \nonumber \\
^{12}{\rm B} \rightarrow ^{12}{\rm B}+^{12}{\rm B}^0\nonumber \\
^{14}{\rm B} \rightarrow ^{14}{\rm B}+^{14}{\rm B}^0 \nonumber \\
^{13}{\rm B} \rightarrow ^{13}{\rm B}+^{14}{\rm B}^* \nonumber \\
^{15}{\rm B} \rightarrow ^{15}{\rm B}+^{16}{\rm B}^* \nonumber \\
^{17}{\rm B} \rightarrow ^{17}{\rm B}+^{18}{\rm B}^* \nonumber \\
^{15}{\rm C} \rightarrow ^{15}{\rm C}+^{15}{\rm C}^0 \nonumber \\
^{14}{\rm C} \rightarrow ^{14}{\rm C}+^{15}{\rm C}^* \nonumber \\
^{16}{\rm C} \rightarrow ^{16}{\rm C}+^{16}{\rm C}^0+^{17}{\rm C}^* \nonumber \\
^{17}{\rm C} \rightarrow ^{17}{\rm C}+^{17}{\rm C}^0 \nonumber \\
^{18}{\rm C} \rightarrow ^{18}{\rm C}+^{19}{\rm C}^* 
\end{eqnarray}

In our calculation excited states are accounted for by multiplying
the ground state densities by the $R$ factors such as those in Ref.
\cite{Natowitz23}, see also the Supplemental material to \cite{Roepke21}. These are defined as:

\begin{figure}
 \begin{tabular}{c}
 \includegraphics[width=0.45\textwidth]{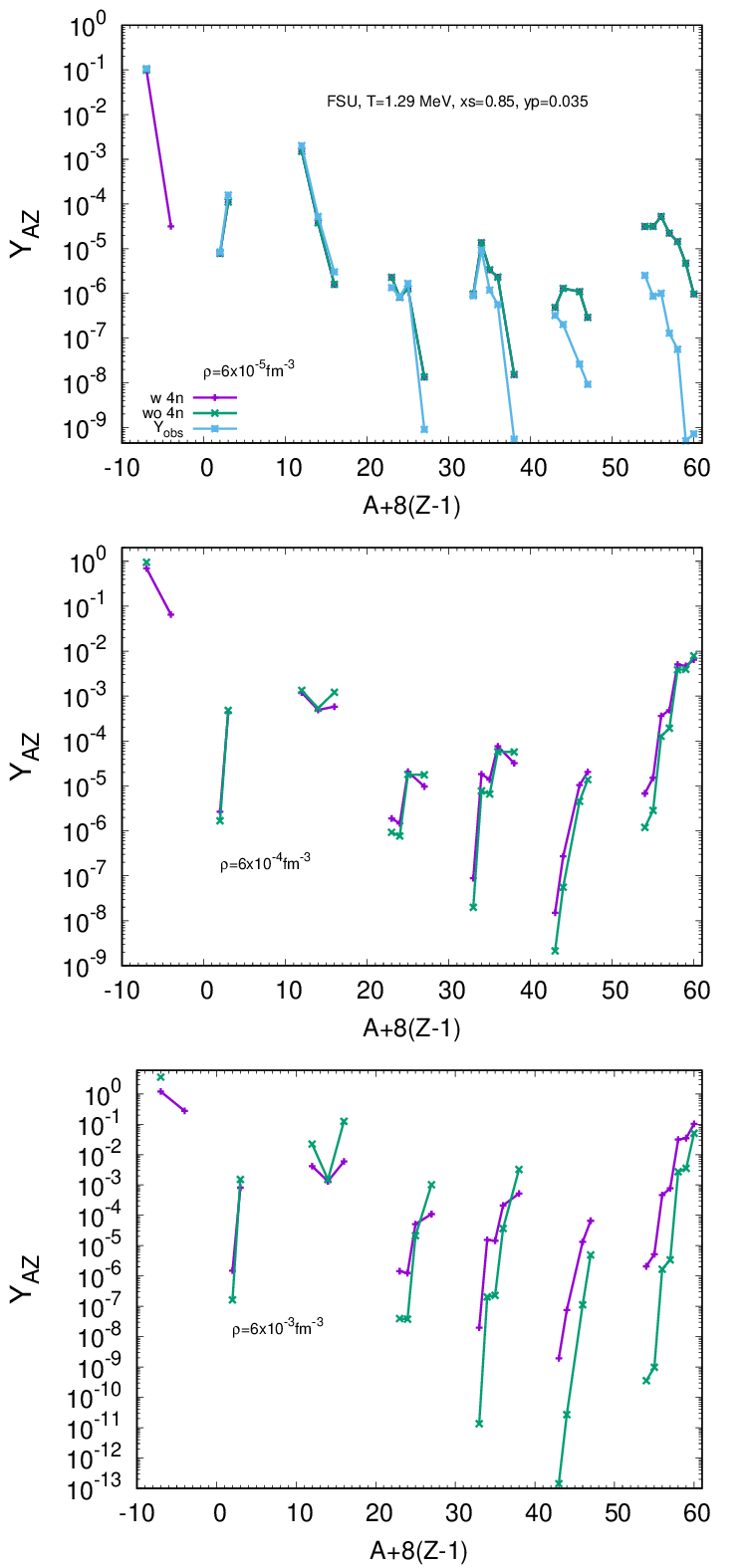}
  \end{tabular}
  \caption{Comparison between calculated and effective yields as a function of the Lestone parameter, $A+8(Z-1)$, for the RMF FSU model with (dark blue) and without (green) the tetraneutron, taking $y_p=0.035$, and $T=1.29$ MeV. Also shown are the observed yields (light blue). Each panel corresponds to a different density. The binding energy of $4n$ is taken to be $-2.37$ MeV, and the scalar cluster--meson coupling is equal to $x_s=0.85$.} 
\label{fig2}
\end{figure}

\begin{figure}
 \begin{tabular}{c}
 \includegraphics[width=0.45\textwidth]{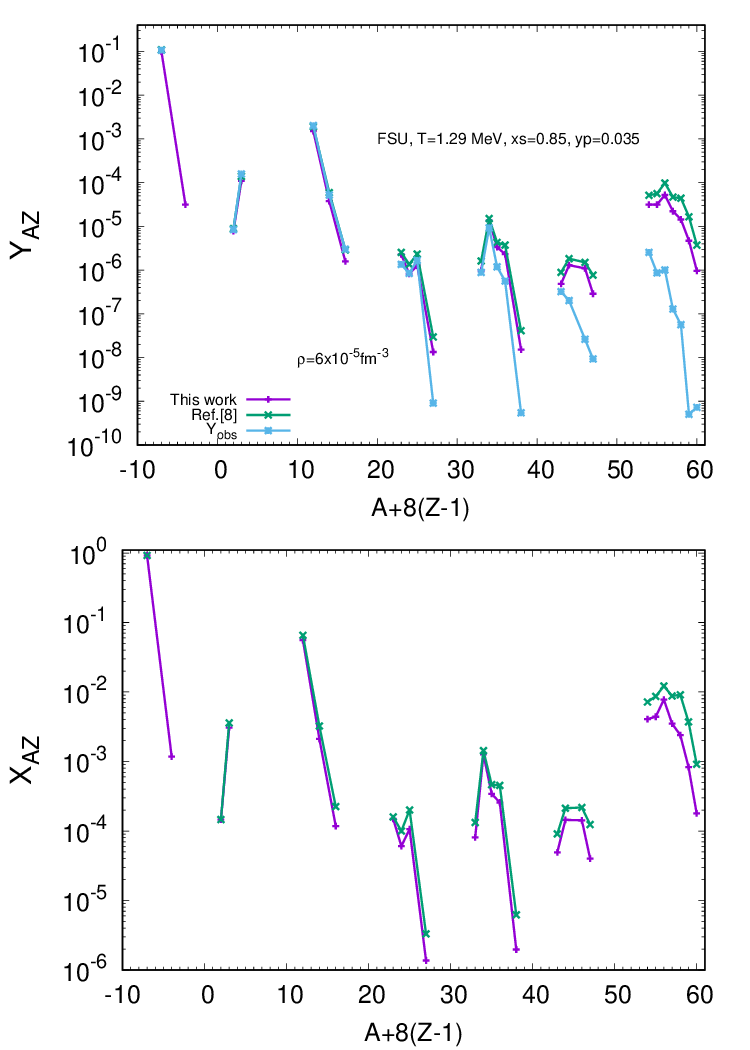}
  \end{tabular}
  \caption{Effective yields (top) and mass fractions (bottom) as a function of the Lestone parameter, $A+8(Z-1)$, for the RMF FSU model (dark blue), taking $y_p=0.035$, and $T=1.29$ MeV, and a density of $6\times 10^{-5}$ fm$^{-3}$. Also shown are the observed yields (light blue), and the calculation taken from \cite{Natowitz23} (green) (see the last two columns of Tables I and II therein).} 
\label{fig3}
\end{figure}

\begin{figure*}
 \begin{tabular}{c}
 \includegraphics[width=0.85\textwidth]{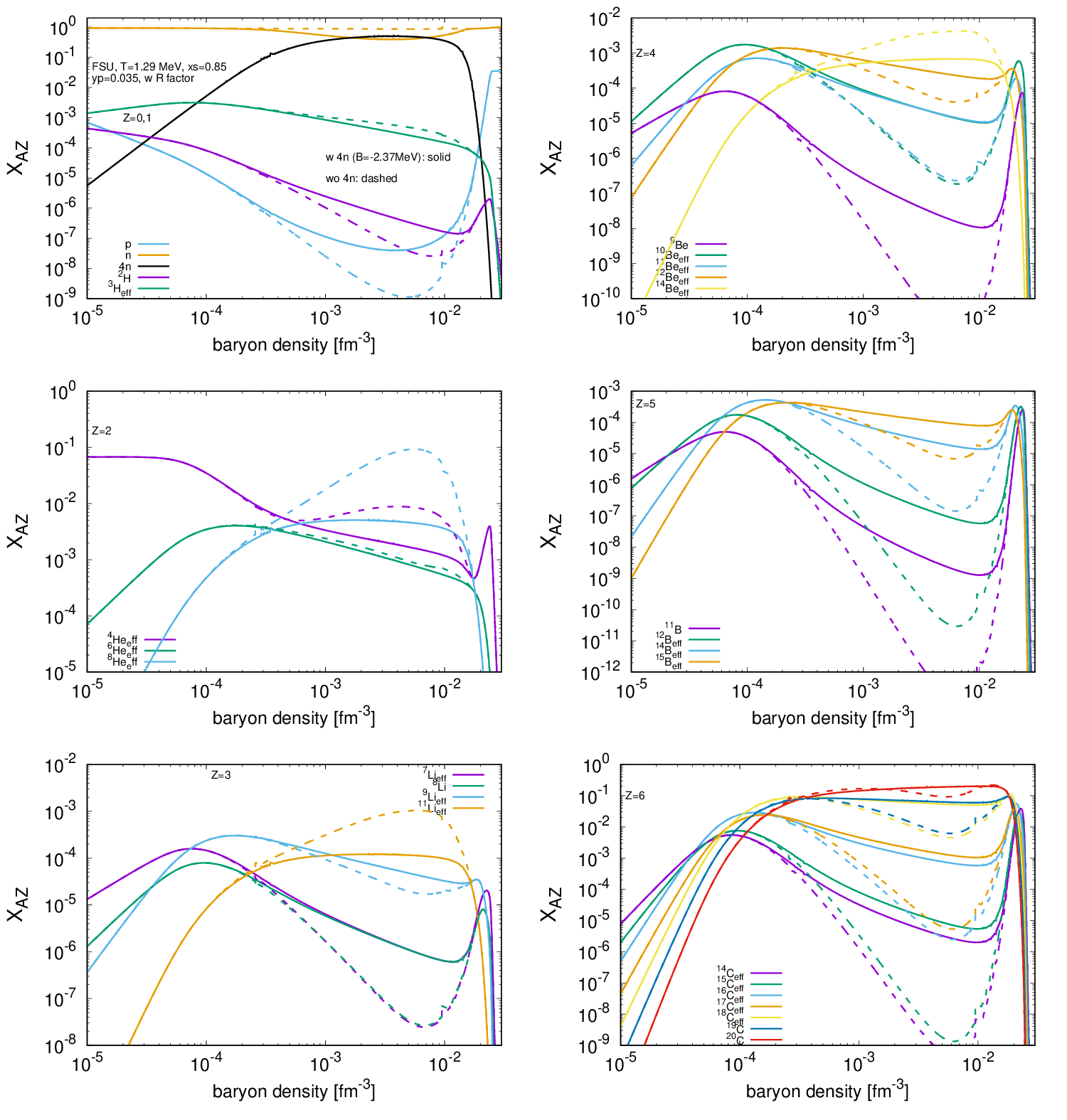}
  \end{tabular}
  \caption{Mass fractions as a function of the density for the RMF FSU model with (solid) and without (dashed) the tetraneutron, taking $y_p=0.035$, and $T=1.29$MeV. Each panel corresponds to a different element. The binding energy of $4n$  resonance is taken to be $-2.37$MeV, and the scalar cluster--meson coupling is equal to $x_s=0.85$.} 
\label{fig4}
\end{figure*}

\begin{eqnarray}
R_i&=&1+\sum_i g_i/g_{A,Z}\exp(-E_i/T) \, ,
\end{eqnarray}
where $i$ represents the excited states included. $E_i, g_i, g_{A,Z}$ can be found in Tables IV - VIII of Ref.~\cite{Natowitz23}.

We first investigate whether the account of the 4$n$ resonance will influence the yields which are observed in ternary fission. We use the data observed from ternary fission of $^{241}$Pu \cite{Koes1,Koes2} and determine the freeze-out parameter values for temperature $T$ and baryon density $n$. 
Following the approaches of Refs.~\cite{Sara,Roepke21,Natowitz23}, the fit metric  given there is applied  to extract optimum values for the freeze-out parameters at which ternary fission occurs. 
A calculation without the $4n$ resonance is compared with a calculation which takes $4n$ resonances into account.  
In Fig.~\ref{fig1}, fits with and without the inclusion of the $4n$ cluster and two different $4n$ binding energies are depicted for various groups of elements. 
We see that the optimum density, where the fit metric as function of the baryon density takes the minimum, is unchanged if the $4n$ resonance is included, and decreases slightly as $Z$ increases.
However, the variation is small, extending over a range of densities of $4 \times 10^{-5}$ to $8 \times 10^{-5}$ fm$^{-3}$. The intermediate value, $6 \times 10^{-5}$ fm$^{-3}$, has been considered in the next two figures.

In Figure~\ref{fig2}, the effective yields of the light isotopes  with inclusion of the $4n$ resonance (dark blue) and without inclusion of $4n$ (green) are plotted against the Lestone parameter, $A+8(Z-1)$ \cite{Lestone08}.
For comparison, in the top panel, the experimental yields (light blue) are also shown. The yields are calculated for three fixed densities: $6\times 10^{-5}$ (top), $6\times 10^{-4}$ (middle), and $6\times 10^{-3}$ (bottom) fm$^{-3}$, at a temperature of 1.29 MeV and a proton fraction of 0.035.  For the smallest density considered, there is no difference between the calculation with and without the $4n$ resonance, as this cluster has negligible yields, for the thermodynamic conditions considered. 
Note that at density $6\times 10^{-5}$  fm$^{-3}$ the calculated yields for $Z > 3$ elements are greater than observed from ternary fission of $^{241}$Pu \cite{Koes1,Koes2}.
This has been attributed to nucleation time and/or finite size effects in our previous work  \cite{Natowitz23,Sara}.
For the other two densities, as the Lestone parameter increases, i.e. for He and heavier elements, the calculated difference between  $4n$ inclusion and no $4n$ inclusion starts to be noticeable.

\begin{figure*}
 \begin{tabular}{c}
 \includegraphics[width=0.85\textwidth]{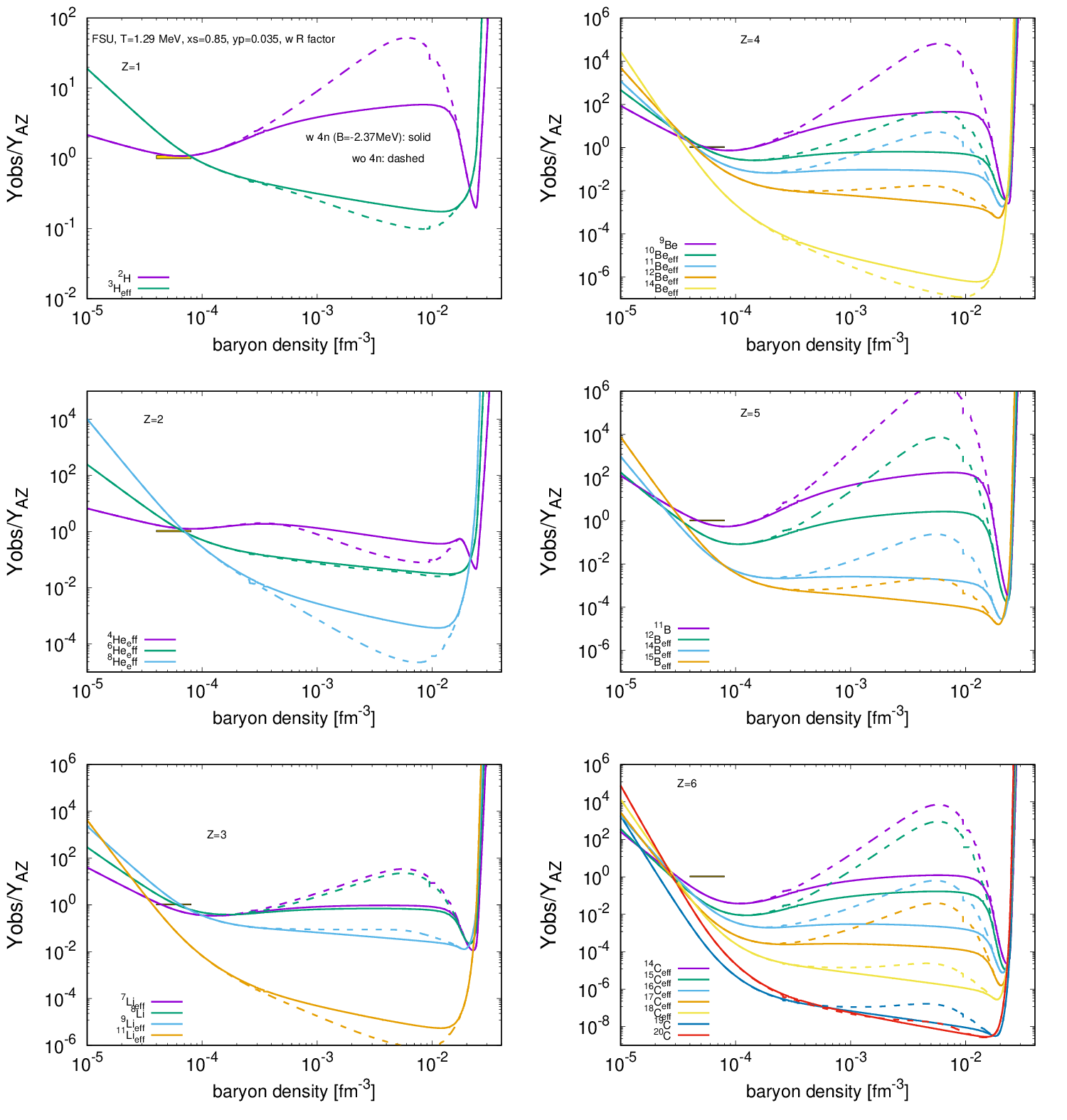}
  \end{tabular}
  \caption{Ratio of the observed yields to the effective yields as a function of the density for the RMF FSU model with (solid) and without (dashed) the tetraneutron, taking $y_p=0.035$, and $T=1.29$ MeV. Each panel corresponds to a different element. The binding energy of $4n$ is taken to be $-2.37$ MeV, and the scalar cluster--meson coupling is equal to $x_s=0.85$. The horizontal line indicates $Y_{obs}/Y_{AZ}=1$ in the density range $[4\times 10^{-5},8\times 10^{-5}]$ fm$^{-3}$.} 
\label{fig5}
\end{figure*}

In Figure~\ref{fig3}, the effective yields (top) and the mass fractions (bottom) are again plotted against the Lestone parameter, for a density of $6\times 10^{-5}$ fm$^{-3}$ and the same temperature and proton fraction as in Fig.~\ref{fig2}. In the top panel, a comparison with results of a different calculation taken from Ref.~\cite{Natowitz23} (green) (see the last two columns of Tables I and II therein) is shown. Both calculations deviate from the observed  yields, especially for Li and above.

In Figure~\ref{fig4}, we present calculated mass fractions as a function
of density with and without the inclusion of tetraneutrons. The
temperature and proton fraction used are again those determined for
ternary fission of $^{241}$Pu. At the onset and dissolution of the clusters, the effect of $4n$ is negligible. Only in the region of densities from $\sim 10^{-4}$ to $\sim 10^{-2}$ fm$^{-3}$, can we see that the $4n$ affects the abundances of other clusters. 
Looking at the top left panel of Figure~\ref{fig4}, we see that the $4n$ can even become the most abundant cluster. This may not be that surprising as we are in a very neutron rich environment ($y_p=0.035$). 
For the lightest elements in the two top left panels, $Z=1,2$,the abundances of the most neutron-rich clusters decrease in the calculation including  $4n$ since neutrons are being consumed by the $4n$. 
For the larger $Z$ elements, abundances increase when the $4n$ is present.
We also observe that, looking at each element individually, the most neutron-rich isotopes are the most abundant, irrespective of the $4n$ being in the system or not. 
Close to the dissolution, we observe that for some clusters, e.g.,  $^2$H or $^4$He, there is a peak, a sudden increase of the mass fraction. 
This is explained by the fact that some of the competing neutron rich clusters, like $4n$ or $^3$H, decrease in abundance making more  neutrons available for other clusters. 
A similar behavior was also observed in Ref.~\cite{Pais18}, Fig.~5.

In Figure~\ref{fig5} the
ratios of experimentally observed yields from the ternary fission of
$^{241}$Pu to the calculated yields with and without the tetraneutrons
are plotted against the density. 
The horizontal line indicates the ratio equal to 1, i.e. where the experimental values are equal to the theoretical ones, in the range of densities that include the minima of the fit metric plot, Fig.~\ref{fig1}, $4\times 10^{-5}$ to $8\times 10^{-5}$ fm$^{-3}$. 

Examination of Figures \ref{fig4} and \ref{fig5}  reveals that the presence
of tetraneutrons with the properties considered in this work should have no significant effect in the low density region
previously extracted from the ternary fission data. This region is
signaled in Figure \ref{fig5} by the ratios equal to 1.

\begin{figure}
 \begin{tabular}{c}
 \includegraphics[width=0.45\textwidth]{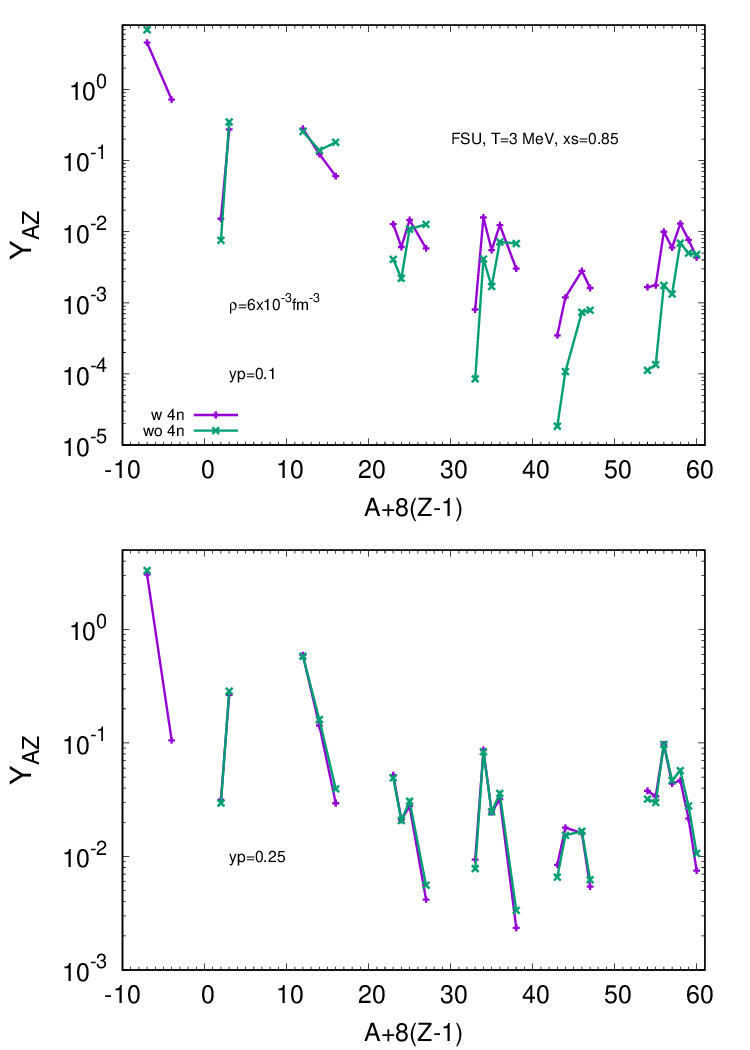}
  \end{tabular}
  \caption{Effective yields as a function of the Lestone parameter, $A+8(Z-1)$, for the RMF FSU model with (dark blue) and without (green) the tetraneutron, taking $y_p=0.1$ (top) and $y_p=0.25$ (bottom), and $T=3$ MeV. The binding energy of $4n$ is taken to be $-2.37$ MeV, and the scalar cluster meson coupling is equal to $x_s=0.85$.} 
\label{fig6}
\end{figure}

So far we have considered a temperature of 1.29 MeV, and a very low proton fraction of 0.035,  conditions attained in low-energy ternary fission experiments. 
In the last figure of this paper, Figure~\ref{fig6}, we show how the yields calculated with and without inclusion of tetraneutrons at a density of  $6\times 10^{-3}$ fm$^{-3}$ vary  when larger proton fractions, 0.1 and 0.25, and a higher temperature, 3 MeV are considered.
We see that the yields increase with respect to the ones calculated at $T=1.29$ MeV. We also observe differences   between the results of the calculation with and without inclusion of the $4n$ resonance, especially for the lower proton fraction case.  
This difference becomes larger as $Z$ increases. 
These observed effects at higher density and temperature suggest that it may be possible to carry out tests of tetraneutron effects in nuclear reactions capable of probing denser, higher temperature matter than that reached in the neck region of spontaneous or neutron induced ternary fission events. 
One such case might be mid-peripheral heavy ion collisions in which the emission of particles and fragments from the contact region, a process similar to ternary fission \cite{Abdurrahman24}, occurs.

\section{Conclusions}

A relativistic mean-field (RMF) formalism, with  in-medium effects included  has been used to carry out a theoretical analysis of the effect of including a tetraneutron resonant state with $E_{4n}=2.37\pm 0.38$(stat)$\pm 0.44$(sys) MeV on fragment formation in low density, low temperature, neutron rich matter such as that existing in the neck region of ternary fissioning nuclei. 
At the density, temperature and proton fraction characterizing the neck region of spontaneous or thermal neutron induced fission no significant difference between yields calculated with and without the inclusion of the $4n$ resonance is observed.
However, for the same temperature and proton fraction, at higher densities strong differences are seen between the yields calculated with and without inclusion of the tetraneutron.  

A calculation at higher density, $6\times 10^{-3}$ fm$^{-3}$, $T=3$ MeV  and $y_p=0.1$ suggests that   selection of suitable reaction dynamics scenarios and bombarding energies could allow laboratory searches for the effects of tetraneutrons over a range of temperatures, densities and proton fractions. A recent study employing very light heavy ions in peripheral collisions may provide some additional support for this conclusion \cite{Muzalevskii23}.

\section*{Acknowledgements}
This work was partially supported by portuguese national funds from FCT (Funda\c c\~ao para a Ci\^encia e a Tecnologia, I.P, Portugal) under projects UIDB/04564/2020 and UIDP/04564/2020, with DOI identifiers 10.54499/UIDB/04564/2020 and 10.54499/UIDP/04564/2020, respectively, and the project 2022.06460.PTDC with the associated DOI identifier 10.54499/2022.06460.PTDC. H.P. acknowledges the grant 2022.03966.CEECIND (FCT, Portugal) with DOI identifier 10.54499/2022.03966.CEECIND/CP1714/CT0004. This work was also supported by the US Department of Energy under Grant No. DE–FG02–93ER40773.

\end{document}